\documentclass[10pt]{article}
\usepackage[margin=1in]{geometry}
\usepackage{titling}
\usepackage{lipsum}
\usepackage{multicol}
\usepackage{graphicx}
\usepackage{draftwatermark}
\usepackage{color,soul}
\usepackage{float}
\SetWatermarkText{DRAFT - DRAFT}
\SetWatermarkScale{3}

\title{\textbf{ PubSqueezer: A Text-Mining Web Tool to Transform Unstructured Documents into Structured Data
}}
\author{Alberto Calderone, Ph.D.\\
		sinnefa@gmail.com\\
		}
\date{\today}

\begin{document}
\maketitle

\begin{abstract}
The amount of scientific papers published every day is daunting and constantly increasing. Keeping up with literature represents a challenge. If one wants to start exploring new topics it is hard to have a big picture without reading lots of articles. Furthermore, as one reads through literature, making mental connections is crucial to ask new questions which might lead to discoveries. In this work, I present a web tool which uses a Text Mining strategy to transform large collections of unstructured biomedical articles into structured data. Generated results give a quick overview on complex topics which can possibly suggest not explicitly reported information. In particular, I show two Data Science analyses. First, I present a literature based rare diseases network build using this tool in the hope that it will help clarify some aspects of these less popular pathologies. Secondly, I show how a literature based analysis conducted with PubSqueezer results allows to describe known facts about SARS-CoV-2. In one sentence, data generated with PubSqueezer make it easy to use scientific literate in any computational analysis such as machine learning, natural language processing etc. \\

Availability: http://www.pubsqueezer.com
\end{abstract}

\begin{multicols}{2}

\section{Introduction}
Scientific literature is the entry point to the understanding of any research topic. While key ideas identification in texts is useful when reading articles to spot important words and phrases, much information is usually scattered in many different articles and often can only be derived by mental connections.

Reading several articles about one topic is an essential and yet time consuming activity aimed to make mental connections and come up with new hypotheses which are not yet explicitly reported in the literature itself. Automatic text analyses can facilitate researcher's mental process by speeding up tasks such as keywords and key phrases identification as well as making cross connections amongst a large number of papers. 

Text mining (TM) is the process of extracting information from documents by identifying text patterns via computational and statistical approaches. Some TM tools such as SciLite \cite{Venkatesan} aim to keywords highlighting. Differently, to support the process of extracting essential information, I developed PubSqueezer (http://www.pubsqueezer.com), a tool which aims at analysing and integrating multiple articles in order to extract not explicitly written information. PubSqueezer aims to transform unstructured collections of documents into a structured format which can be used for literature exploration and Data Science analyses.

\section{Results}

\begin{figure*}
  \includegraphics[width=\textwidth]{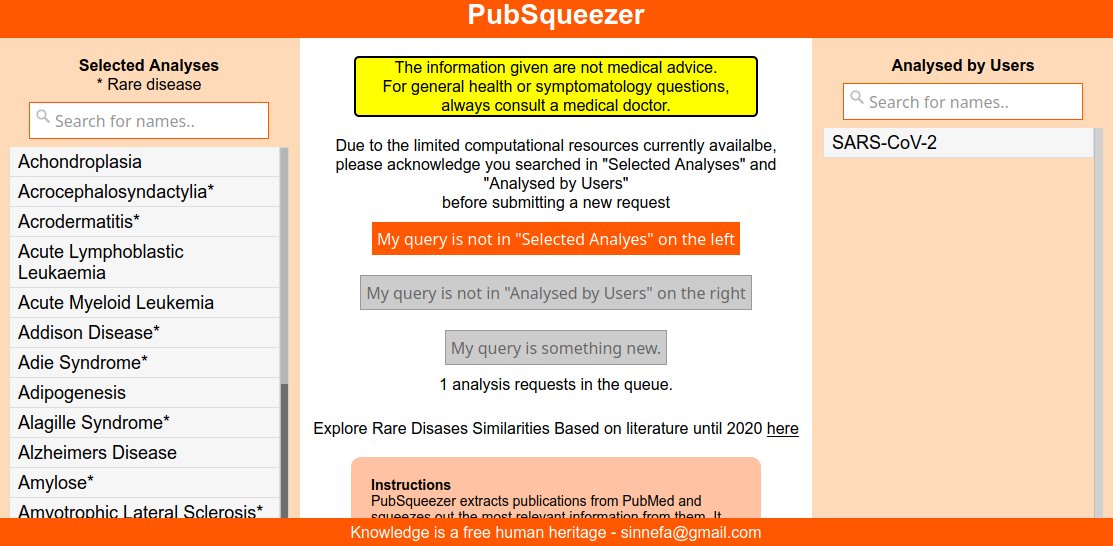}
  \caption{PubSqueezer Homepage}
   \label{figure1}
\end{figure*}

PubSqueezer is a TextMining web engine available at http://www.pubsqueezer.com which directly queries PubMed (https://pubmed.ncbi.nlm.nih.gov/). By downloading publications from PubMed, Pubsqueezer analyses the latest 2,000 (if abstracts available) publications about a give topic to mines gene names, key phrases and words, and ranks them according to their relevance. Due to the limited computational power of the hosting server, the analysis takes some time but each analysis is saved and listed in the homepage. The homepage (Figure \ref{figure1}) provides two lists. The one on the left, is a list of selected queries while the list on the right is a list of users' queries. The more PubSqueezer will be used, the more analyses will be publically available. These two lists can save time to any user searching for something already queried in the past.

\begin{figure*}
  \includegraphics[width=\textwidth]{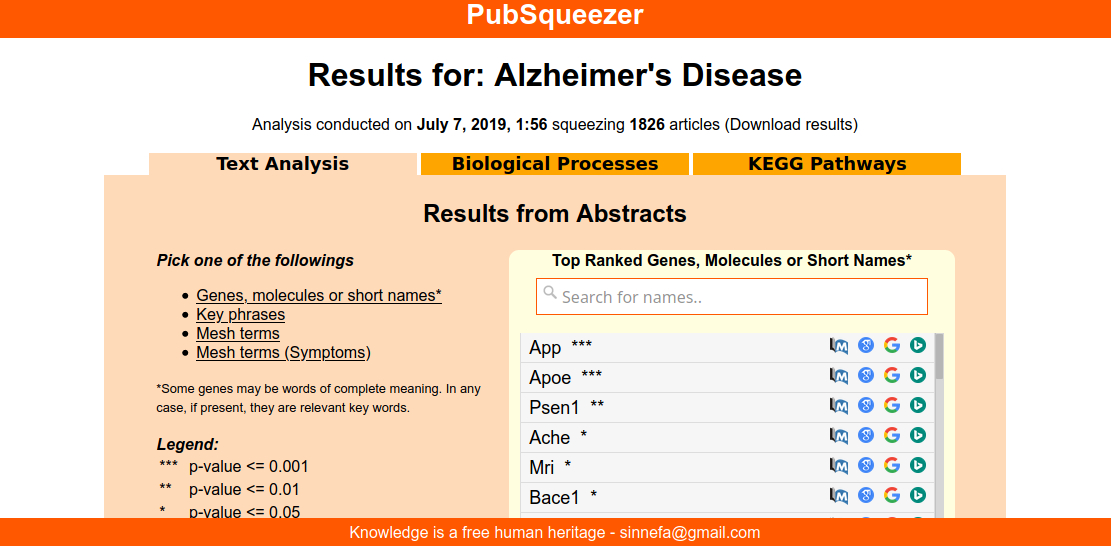}
  \caption{PubSqueezer Results page. You can pick genes, key phrases or terms ranked according to significance score. The three tabs on top allow you to explore more details.}
   \label{figure2}
\end{figure*}

The result page (Figure \ref{figure2}) is divided into three tabs. The first tab contains key terms and phrases. The second and third tabs are lists of biological pathways and processes (Obtained through GO enrichment \cite{go}) which are relevant to the query. Each result is directly linked to PubMed, Google, Google Scholar and Bing to allow the user to directly retrieving external information. Finally, all the lists shown in the three tabs can be downloaded as a zip file for further computational analyses.

\begin{figure*}
  \includegraphics[width=\textwidth]{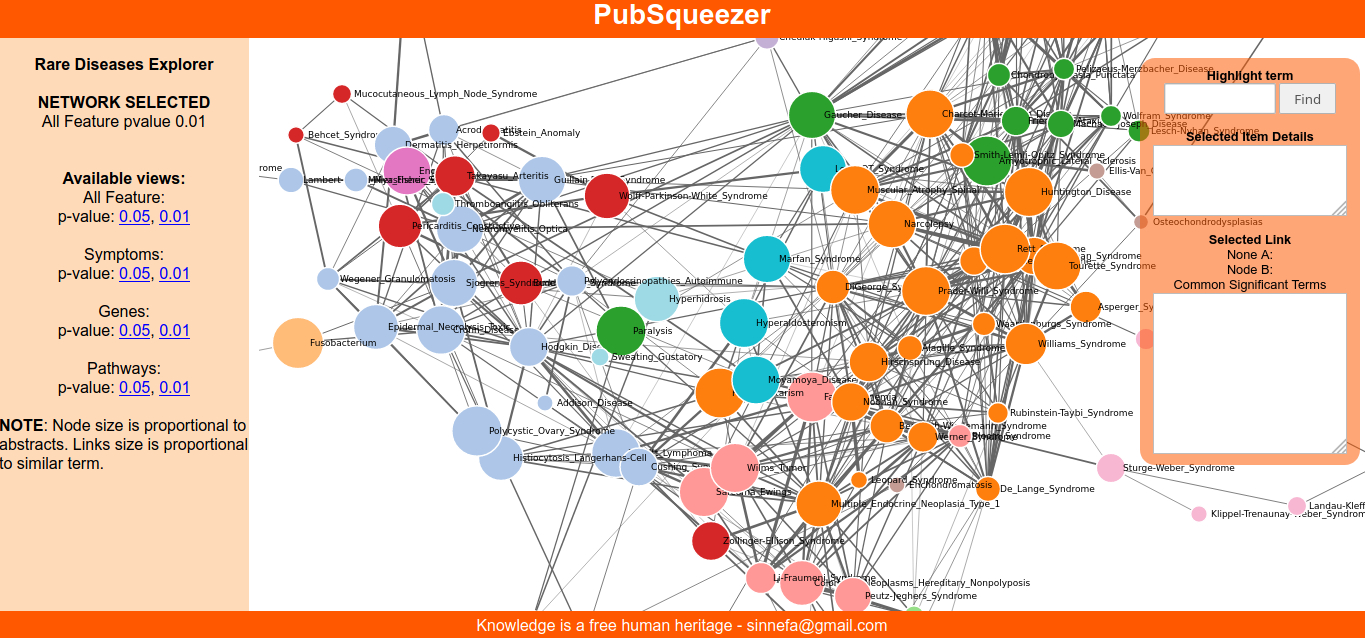}
  \caption{PubSqueezer one of the available Rare Disease Similarity Network. Clicking on nodes and edges you can get extra information. Clicking on diseases you get symptoms, genes atc. while clicking on edges you can see what two pathologies have in common i.e. common pathways, symptoms etc.}
   \label{figure3}
\end{figure*}

PubSquezer also has a section dedicated to Rare Diseases (Figure \ref{figure3}). This section shows similarities among rare diseases according to features varying from symptoms to genes. The network is interactive. Clicking on nodes shows details about a disease while clicking on links shows what two conditions have in common.

\section{Methods}
All PubSqueezer results are obtained through statistic hypothesis testing. All processed texts are preprocessed to remove unnecessary terms, lemmatized and tokenized.

In order to perform a statistical test, a large heterogeneous background set was build. PubSqueezer contains a large background of publications build downloading 2,000 publications for each one of the following keywords: proteomic, proteomics, gene, genetic, genetics, genomic, genomics, DAN, RNA, pathology, syndrome, disease, metabolism, metabolic. These terms were chosen to make the background as varied and as large as possible thus allowing relevant terms to "emerge" through statistical randomization testing.

Upon a user's request, PubSqueezer downloads a set of publications from PubMed and compare its content against the background. As an example, every gene name and keyword which has a significant p-value against the background is considered relevant to the topic. The test is done comparing the query against 2,500 random samples of the same size of the user's set. The number 2,500 was calculated considering an upper bound to obtain a precision level of 0.01 \cite{pvalue}.

All terms considered relevant to the queried topic are ranked according to their p-values. Key phrases are extracted by scanning abstracts with the RAKE algorithm \cite{rake}. Finally, pathways and biological processes are derived from KEGG\cite{kegg} using significant gene names. Each result is directly linked to PubMed, Google, Google Scholar and Bing websites to allow further investigation.

\begin{table*}
\begin{tabular}{cc}
	\includegraphics[height=1.5in]{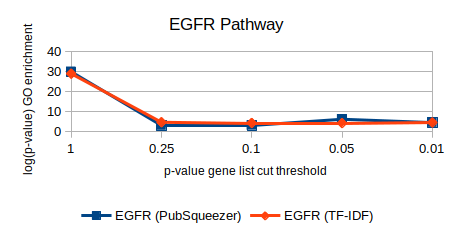} & \includegraphics[height=1.5in]{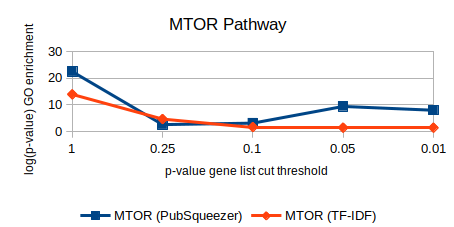}\\
	\includegraphics[height=1.5in]{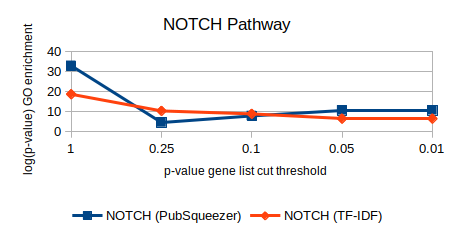} & \includegraphics[height=1.5in]{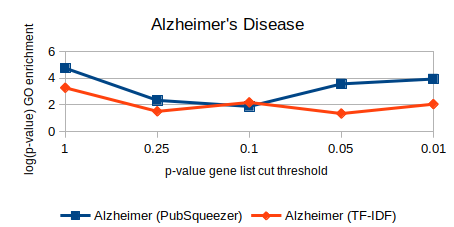}\\
\end{tabular}
\caption{PubSqueezer ranked lists VS. TF-IDF. Comparing p-value obtained by some processes using gene names filtered with different. Overall PbuSqueezer terms ranking seems to me better than TF-IDF.}
\label{table1}
\end{table*}

In order to assess PubSqueezer results quality I tried to use it with 4 different processes: 3 pathways and one pathology. Using the results obtained, I compared the scores assigned by PubSqueezer with those obtained with the classic TF-IDF\cite{tfidf} algorithm. Overall, it seems like PubSqueezer ranking works better than TD-IDF. Taking PubSqueezer and TF-IDF lists and iteratively cutting them at different thresholds one can select top-X results. Using these genes with GO enrichment \cite{go} it is possible to see if the expected results obtain a significant p-value. As shown in the plot-table (Table \ref{table1}) PubSqueezer - with the exception of EGFR - tends to assign better scores (higher in the ranking) to more important genes.

Finally, to test if it is possible to derive not explicitly reported facts from papers using PubSqueezer, I did an analysis on the Alzheimer’s Disease and Parkinson’s Diseases which, other than having disease specific pathways, share common processes which can not be detected by the simple disease specific gene ontology enrichment analysis \cite{calderone}. I performed this analysis removing possible hints for the algorithms – i.e. papers which potentially mention these common processes.

\textbf{Query 1}: Alzheimer NOT Glucose NOT Phosphate NOT Metabolism NOT DNA NOT damage NOT Apoptosis NOT (Cell AND Cycle) NOT Protein NOT Localization NOT Vesicles NOT Trafficking NOT RNA NOT regulation NOT transcription

\textbf{Query 2}: Parkinson NOT Glucose NOT Phosphate NOT Metabolism NOT DNA NOT damage NOT Apoptosis NOT (Cell AND Cycle) NOT Protein NOT Localization NOT Vesicles NOT Trafficking NOT RNA NOT regulation NOT transcription

The following table shows biological processes implicitly recoverable through GO enrichment \cite{GO} from not explicitly reported term both from the two single queries and from their intersections. In other words, articles never talked about any of those arguments and yet, using the gene names ranked by PubSqueezer it is actually possible to recover those processes.

\end{multicols}

\begin{table}[h]
\centering
\begin{tabular}{|l|l|l|l|}
\hline
	\textbf{Common processes} & \textbf{Alzheimer} & \textbf{Parkinson} & \textbf{Intersection}\\
\hline
	Glucose Metabolism & NO & YES & NO\\
\hline
	Phosphate Metabolism & YES & YES & YES\\
\hline
	DNA Damage & NO & NO & NO\\
\hline
	Apoptosis & YES & YES & YES\\
\hline
	Cell Cycle & YES & NO & NO\\
\hline
	Protein Localization Vesicles Trafficking & YES & YES & YES\\
\hline
	RNA Metabolism & YES & YES & NO\\
\hline
	Regulation of Transcription & YES & YES & YES\\
\hline
\end{tabular}
\end{table}

\begin{multicols}{2}

\section{Structured Data Exploitation through Data Science}
Other than using the web interface to explore scientific literature going directly to facts which are somewhat statistically relevant to the query, it is also possible to download results as comma-separated-values CSV files so that they can be processed with Data Science strategies. In this section, I show two examples on how to use such data to perform more complex analyses using computational methods.

\subsection{Rare Diseases: discovering connections despite the lack of direct publications}
The first example is to help rare diseases research by projecting knowledge (literature) from one disease to another. Most rare diseases have few literature which hampers research and the understanding of the conditions themselves. On possibility is to see similar pathologies for which literature is available and use them to speculate on some other less studied but connected condition.

\end{multicols}
\begin{figure}[H]
\centering
  \includegraphics[width=400px]{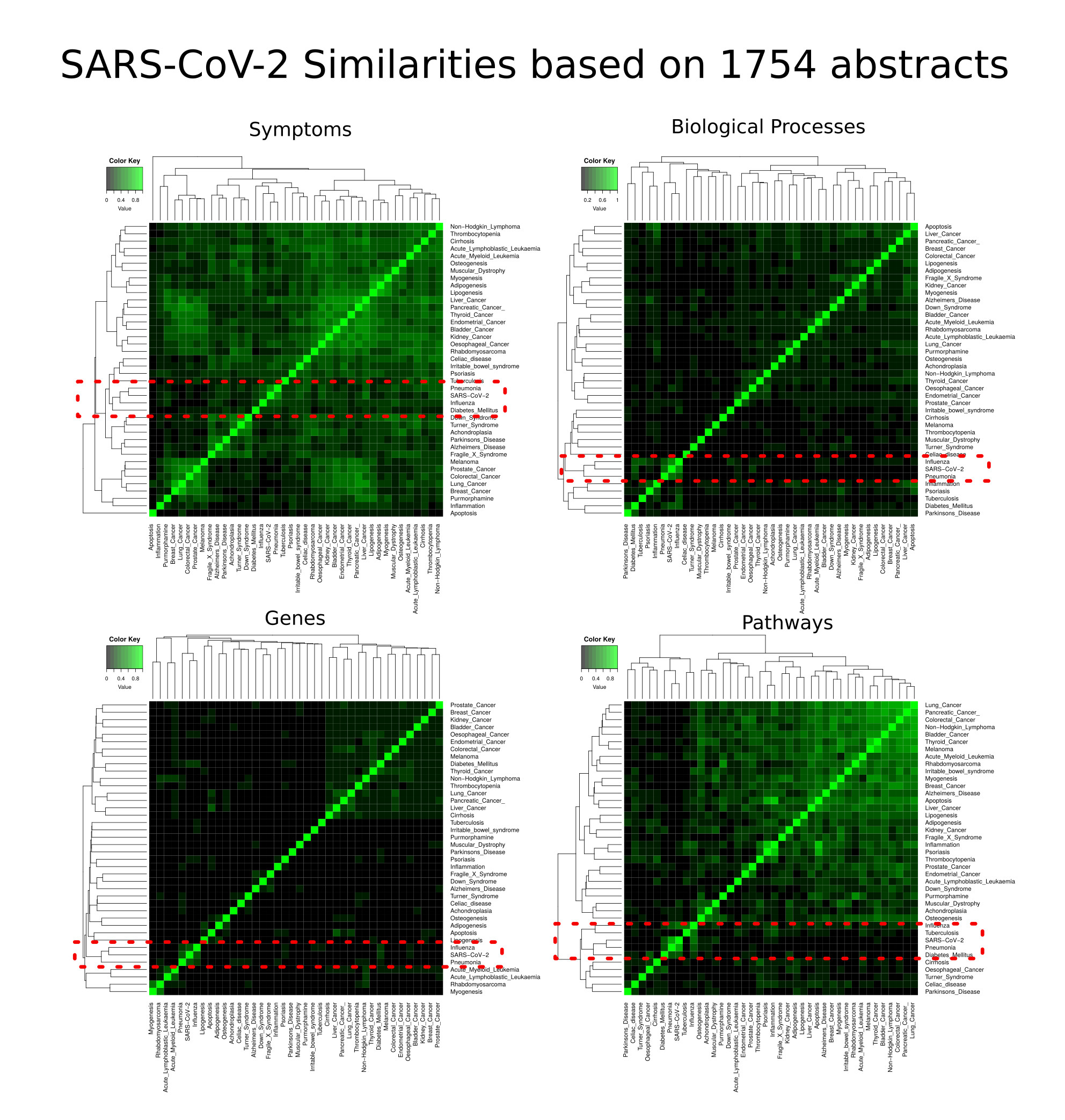}
  \caption{SARS-CoV-2 heat-map comparisons against other conditions. It is also nice to notice similarities among mental conditions and forms of cancer}
   \label{figure4}
\end{figure}
\begin{multicols}{2}

This first strategy aims at building a network of similarities so that the information known about one rare disease can be used to explore similarities among other rare diseases. I used PubSqueezer to query PubMed on thousands of rare diseases. Each query results in CSV's files which can be interpreted as "features" of each condition. Considering these features, one can calculate similarities, let's say what genes some pathology have in common, and link two conditions according to the magnitude of the similarity. In order to compute these similarities I used the classic cosine similarity which tolerates missing components and heterogeneous feature magnitudes. The final result is shown in Figure \ref{figure3}. In the PubSqueezer home page interface you can find several maps to explore similarities on different levels.

\subsection{SARS-CoV-2: using data to get an overview}
A second use case could be the exploration of literature at a glance. When one starts studying a new topic, it is mandatory to dig into the literature to find out what is known so far. Usually one relies on scientific reviews which are build upon many other scientific publications. PubSquezer somewhat make automatic reviews on a topic. In particular, I show how the SARS-CoV-2 query extracts what is known and how we can use these information to make sense out of it, or at least understand what is known so far.
To go beyond the results page, I explore SARS-Cov-2 (Figure \ref{figure4}) using cosine similarity among other conditions. I build similarity heat-maps to highlight what are the cross similarities with SARS-CoV-2. This analysis shows knows facts: SARS-CoV-2 is similar to Pneumonia \cite{pneumonia} and Influenza plus it also has some correlation to diabetes\cite{diab}, especially at the level of biological pathways involved.

\section{Conclusions} While this web tool is still on a very preliminary stage, I show in this article that it might be useful to squeeze multiple articles into structured data that can be further manipulated to derive new information. The hosting server is very limited but hopefully, in the future, I will migrate this to a better server. Results obtained with PubSqueezer are not perfect but already valid as shown in methods, they can help having a quick overview on a topic as well as and, most importantly, transform unstructured data into structured data to promote Data Science analyses. Possibly, the ideas in this draft article can be exploited by others.

\section{Acknowledgements}
I would like thank dr. Elisa Micarelli for supporting my work, listening to my crazy talking and for her suggestions on the preliminary interface. I would thank dr. Andrea Cerquone Perpetuini for listening to my work and for helping me trying different queries and cases and for trying this version of the web interface. Finally, I would like express my sincere gratitude to dr. Elena Santonico for her constant support, reading this draft and for trying this version of the web interface.

\end{multicols}
\end{document}